\documentclass[iop]{emulateapj}

\usepackage{natbib}
\usepackage{longtable}
\def\gtorder{\mathrel{\raise.3ex\hbox{$>$}\mkern-14mu
             \lower0.6ex\hbox{$\sim$}}}
\def\ltorder{\mathrel{\raise.3ex\hbox{$<$}\mkern-14mu
             \lower0.6ex\hbox{$\sim$}}}

\newcommand{\newQSOs}{144 }
\newcommand{\knoQSOs}{25 }
\newcommand{\allQSOs}{169 }
\newcommand{\observed}{845 }

\shorttitle{The Magellanic Quasars Survey. II. Confirmation of \newQSOs New AGNs Behind the Southern Edge of the LMC}
\shortauthors{Koz{\l}owski et al.}

\begin{document}

\title{The Magellanic Quasars Survey. II. Confirmation of \newQSOs New Active Galactic Nuclei Behind the Southern Edge of the Large Magellanic Cloud}

\author{Szymon~Koz{\l}owski\altaffilmark{1},
C.~S.~Kochanek\altaffilmark{2,3},
A.~M.~Jacyszyn\altaffilmark{1},
A.~Udalski\altaffilmark{1},
M.~K.~Szyma{\'n}ski\altaffilmark{1},
R.~Poleski\altaffilmark{1}, 
M.~Kubiak\altaffilmark{1},
I.~Soszy{\'n}ski\altaffilmark{1},
G.~Pietrzy{\'n}ski\altaffilmark{1,4},
{\L}.~Wyrzykowski\altaffilmark{1,5},
K.~Ulaczyk\altaffilmark{1}, and
P.~Pietrukowicz\altaffilmark{1}
}

\altaffiltext{1}{Warsaw University Observatory, Al. Ujazdowskie 4, 00-478 Warszawa, Poland; e-mail: simkoz@astrouw.edu.pl}
\altaffiltext{2}{Department of Astronomy, The Ohio State University, 140 West 18th Avenue, Columbus, OH
43210, USA}
\altaffiltext{3}{The Center for Cosmology and Astroparticle Physics, The Ohio State University,
191 West Woodruff Avenue, Columbus, OH 43210, USA}
\altaffiltext{4}{Departamento de Fisica, Universidad de Concepci{\'o}n, Casilla 160-C, Concepci{\'o}n, Chile}
\altaffiltext{5}{Institute of Astronomy, University of Cambridge, Madingley Road, Cambridge CB3 0HA, UK}

\begin{abstract}
We quadruple the number of quasars known behind the Large Magellanic 
Cloud (LMC) from 55 (42 in the LMC fields of the third phase of the Optical Gravitational Lensing Experiment (OGLE)) to 200 by 
spectroscopically confirming \allQSOs (\newQSOs new) quasars from a sample of \observed observed
candidates in four $\sim$3 deg$^2$ Anglo-Australian Telescope/AAOmega fields south of the LMC center.  
The candidates were selected based on their {\it Spitzer} mid-infrared colors, 
X-ray emission, and/or optical variability properties in the database 
of the OGLE microlensing survey.  
The contaminating sources can be divided into 115 young stellar objects (YSOs), 
17 planetary nebulae (PNe), 39 Be and 24 blue stars, 68 red stars,
and 12 objects classed as either YSO/PN or blue star/YSO.
There are also 402 targets with either featureless spectra or too 
low signal-to-noise ratio for source classification. 
Our quasar sample is 50\% (30\%) complete at $I=18.6$~mag (19.3~mag).
The newly discovered active galactic nuclei (AGNs) provide many additional reference points for proper motion 
studies of the LMC, and the sample includes 10 bright AGNs ($I<18$ mag) 
potentially suitable for absorption line studies.  Their primary use, however, is 
for detailed studies of quasar variability, as they all have long-term,
high cadence, continuously growing light curves from the microlensing 
surveys of the LMC.  Completing the existing Magellanic Quasars Survey fields in the LMC and 
Small Magellanic Cloud should yield a sample of $\sim$700 well-monitored AGNs, and expanding 
it to the larger regions covered by the OGLE-IV survey should yield 
a sample of $\sim$3600 AGNs.
\end{abstract}



\section{Introduction}

\cite{2009ApJ...698..895K}, \cite{2010ApJ...708..927K}, and \cite{2010ApJ...721.1014M} introduced, developed,
and applied a new approach to quantitatively modeling the variability of individual quasars, describing the
light curves as damped random walks (DRW) characterized by two parameters, a timescale and an amplitude.  These
parameters are then found to be correlated with the physical properties of the quasar -- wavelength, luminosity,
and black hole mass.  Unfortunately, accurately measuring the parameters of individual quasars requires
long ($\gtrsim$decade) well-sampled light curves that are generally not available (see \citealt{2011ApJ...728...26M}).  While this may ultimately
be solved by projects such as the Panoramic Survey Telescope and Rapid Response System (PanSTARRS; \citealt{2002SPIE.4836..154K}) 
and the Large Synoptic Survey Telescope (LSST; \citealt{2008arXiv0805.2366I}), their cadences are not ideal and it still 
requires a decade of operations to accumulate the light curves.  This makes it challenging to expand
on this promising new approach to probing quasar physics.

There are, however, quasars\footnote{Quasars, AGNs and QSOs will be used interchangeably throughout this paper.} 
with almost two decades of well-sampled light curves -- the quasars lying in
the regions surveyed by microlensing projects such as the Optical Gravitational Lensing 
Experiment\footnote{\tt http://ogle.astrouw.edu.pl/} (OGLE; e.g., \citealt{2008AcA....58...69U}),
MACHO (e.g., \citealt{2000ApJ...542..281A}), and the Microlensing Observations in Astrophysics (MOA; e.g., \citealt{2002MNRAS.330..137N}).  
For example, the combined phases of the OGLE survey (OGLE-II to OGLE-IV) span 15 years with up to 1000 epochs in $I$-band and a few dozen in $V$-band
for approximately 40 million sources with $I<21$~mag in the Magellanic Clouds, and the light curves
continue to be extended by the OGLE-IV survey.

The challenge is identifying the rare quasars in these
dense stellar fields.  A combination of serendipity (e.g., \citealt{1994PASP..106..843S,2003IAUC.8258....1C}), 
X-ray (\citealt{1997AJ....114.2353C,2002ApJ...569L..15D,2003AJ....126..734D}), 
and crude variability measures (\citealt{2003AJ....125.1330D,2003AJ....125....1G,2005A&A...442..495D})
had succeeded in identifying only 55 and 28 quasars in the LMC and SMC, respectively, as of 2010.  In
\cite{2009ApJ...701..508K}, we showed that the mid-IR selection techniques developed for extragalactic
surveys (\citealt{2005ApJ...631..163S}) could also be applied to dense stellar regions, albeit with some contamination from
the relatively rare dusty stars (e.g., young stellar objects (YSOs), planetary nebulae (PNe), etc.).  In  \cite{2010ApJ...708..927K}, we showed
that the DRW variability parameters of quasars differed from those of variable stars, providing a more quantitative
approach to variability-selecting quasars that has been expanded on in \cite{2011ApJ...728...26M} and \cite{2011AJ....141...93B}.  In a single
test observation under poor conditions, \cite{2011ApJS..194...22K} doubled the number of known quasars
behind the SMC from 28 to 57.

Apart from studying variability, quasars in these fields are important tools for deriving more precise proper motions 
of the LMC, SMC and, potentially, the Galactic Bulge  using the {\it Hubble Space Telescope (HST)} or, in the future, 
the {\it James Webb Space Telescope (JWST)}.  \cite{2006ApJ...638..772K} and \cite{2008AJ....135.1024P} based their
LMC proper motions on 21 quasars.  Further improvements require better measuring the internal motions of the
Clouds, and this requires larger numbers of quasars well-distributed over the LMC disk.  Based on
\cite{2008AJ....135.1024P}, the newly confirmed LMC quasars we discuss here could be used to reduce
the proper motion uncertainties by a factor of three, since our new quasars are distributed over a large
$6 \times 4$ deg$^2$ area of the LMC disk.  We will also identify any bright quasars 
suitable for absorption studies of the LMC's interstellar medium (ISM), as we expect 1, 10, and 74 
quasars brighter than $I<17$, 18, and 19~mag in the four observed fields.

The goal of our Magellanic Quasars Survey (MQS) is to identify and characterize all these quasars, where we
estimate a final sample of roughly $700$ quasars in a complete survey of the OGLE-III project regions. 
Spectroscopy is essential not only for confirmation of the candidates but also to determine the 
luminosity, rest-wavelength and black hole masses (through the emission line widths) of the quasars,
so that the variability parameters can be related to other physical properties.  Relatively large
numbers of quasars will be needed to explore this (minimally) three dimensional parameter space
quantitatively.  While the sample will be smaller than the $\sim 9000$ quasars we considered in
\cite{2010ApJ...721.1014M}, the longer, more densely sampled light curves should let us examine
the physical correlations with less ``blurring'' from the noisy parameter estimates obtained from
the sparsely sampled Sloan Digital Sky Survey (SDSS) light curves, making it far easier to determine the intrinsic parameter
distributions. In this paper, we report our preliminary results for the LMC, sadly reduced in scope
by hurricane Yasi. In Section~\ref{sec:agnsel}, we describe the selection methods leading to a sample
of 2434 quasar candidates behind LMC. Section~\ref{sec:data} describes the observation plan, the observations,
and the data analysis.  Section~\ref{sec:newquasars} introduces the new quasars and 
Section~\ref{sec:contamination} discusses the contaminating sources. 
We discuss the results in Section~\ref{sec:discussion} and summarize the paper in 
Section~\ref{sec:summary}.


\section{AGN candidate selection}
\label{sec:agnsel}

We selected targets based on their mid-IR, X-ray and/or photometric variability characteristics and we describe each method in detail below.
The observations were designed for AAOmega, a 400-fiber spectrograph mounted on the 3.9 m Anglo-Australian Telescope 
(AAT; \citealt{2006SPIE.6269E..14S}).
The AAOmega instrument has many more fibers than the expected number of quasars in its field of view, 
so we kept candidates satisfying any of the selection criteria, thereby aiming for higher completeness at the 
price of higher contamination.  There is no point in having empty fibers even if it means observing
more contaminating stars.

\subsection{Mid-IR-selected AGN Candidates}
\label{sec:mIRsel}

\cite{2005ApJ...631..163S} showed that active galactic nuclei (AGNs) have redder mid-IR colors than galaxies and stars, and that they form 
a distinctive group in the mid-IR color-color plane 
(see \citealt{2004ApJS..154..166L} and \citealt{2004ApJS..154...48E}, and also \citealt{2010ApJ...713..970A}, \citealt{2008ApJ...679.1040G}, and 
\citealt{2010ApJ...716..530K} for the mid-IR colors of X-ray and variability selected AGNs).  The red mid-IR colors
of a typical AGN are not due to dust emission, so mid-IR-selected AGNs are generally normal $z\gtorder 1$
broad-line quasars (see \citealt{2010ApJ...713..970A}, \citealt{2011ApJ...728...56A}).  There is also
no significant tendency for them to be Type~II quasars.   This is particularly true with the addition of
a magnitude limit for the optical spectroscopy, since at most redshifts the emission from the disk is 
required to render the source bright enough to be observed.    

In \cite{2009ApJ...701..508K}, we used this approach to identify quasars in the LMC and SMC from the 
{\it Spitzer/IRAC} four-band photometry (i.e., 3.6, 4.5, 5.8, and 8.0$\micron$) of the Surveying the Agents of a Galaxy's Evolution project (SAGE; \citealt{2006AJ....132.2268M}). 
The 5000 AGN candidates were divided into several categories. First, all the objects must be located inside the ``Stern AGN wedge'' (\citealt{2005ApJ...631..163S}).
The locus of cool blackbodies passes through the wedge so the quasar sample is divided into group ``A'' if they are far from the blackbody locus, 
and ``B'' if they can be contaminated with cold stars. 
Second, the mid-IR color-magnitude diagram (CMD) was divided into regions heavily contaminated by YSOs and one that should principally contain quasars, 
based on mid-IR properties of YSOs from \cite{2008AJ....136...18W} and 
the mid-IR data from the 9 deg$^2$ of the Spitzer Deep Wide-Field Survey (SDWFS; \citealt{2009ApJ...701..428A}) 
of spectroscopically confirmed quasars from the AGN and Galaxy Evolution Survey (AGES; \citealt{2011arXiv1110.4371K}).
Finally, the candidates were classed as ``a'' if they had the optical-to-mid-IR colors of quasars in AGES and 
``b'' if they did not.

The most interesting candidates are classed as QSO-Aa (quasar region, away from blackbody locus with QSO optical-to-mid-IR colors), 
but given the large number of AAOmega fibers, 
we targeted all classes of objects, excluding only those marked as ``outside'' (of the OGLE-III fields) or ``faint'' (not detected on the OGLE-III template images).

\subsection{X-Ray-selected AGN Candidates}

We searched for the OGLE-III counterparts to the 758 X-ray sources distributed over $10\times10$ deg$^2$ of the LMC from 
\cite{1999A&AS..139..277H}.  The typical positional uncertainly for these X-ray sources is several arcseconds, so we
cannot effectively target them without some additional selection criterion given the stellar densities.  As a second
criterion we chose the most optically variable source within $3\sigma$ of the X-ray position.  In this case, variability
simply means that the light curve is inconsistent with a constant flux, without any of the more quantitative variability
selection criteria we discuss in the next section.   This led to a sample of 205 X-ray candidates, of which 160 were also
mid-IR-selected candidates and 99 satisfy the variability cuts presented in Section~\ref{sec:varsel}.

\subsection{Variability-selected AGN Candidates}
\label{sec:varsel}

The OGLE-III database (\citealt{2008AcA....58...69U,2008AcA....58...89U}) contains the data for 9 years (2001--2009) of 
continuously monitoring $\sim$35 million objects toward the LMC. 
We used this database to search for variable objects that are likely quasars based on the DRW model of their light curves (e.g., \citealt{2009ApJ...698..895K,2010ApJ...708..927K,2010ApJ...721.1014M}).

We prepared and analyzed the light curves as described in \cite{2010ApJ...708..927K}. We  use the timescale $\tau$, scaled amplitude $\hat{\sigma}^2=2\sigma^2/\tau$,
and the likelihood ratio $\ln L_{\rm best}/\ln L_{\rm noise}$ between the best fitting model and a white noise model corresponding to simply expanding the photometric errors.
We also fit a power-law structure function (SF) and estimated its slope $\gamma$ between 30 days and 2 years, and an amplitude $A$ defined by the magnitude difference 
between the first and third quartile of the sorted light curve.

To remove unwanted variable sources, we use the following cuts:
{\it Cut 1.} The average light curve magnitude is $I<19.5$~mag;
{\it Cut 2.} $\ln L_{\rm best}>\ln L_{\rm noise}+2$;
{\it Cut 3.} The SF slope $0.1<\gamma<0.9$ (e.g., see \citealt{2010ApJ...714.1194S});
{\it Cut 4.} The $I$-band amplitude $A<0.4$~mag (this removes large amplitude variable stars);
\cite{2010ApJ...721.1014M} report that some of $\sim$9000 SDSS quasars occupy areas outside the $\tau$--$\hat{\sigma}$ cut of \cite{2010ApJ...708..927K}, 
therefore we decided to loosen the cut on the $\tau$ parameter and to ignore $\hat{\sigma}$ parameter, so {\it Cut 5} is $1<\log(\tau)<5$.
When simply applied to the OGLE light curve database, this process yields $\sim 24,000$ candidates.  The vast 
majority are due to two known systematic issues.  First, bright variable stars typically generate several
fake, fainter variable stars in their wings, and these ``ghost'' variables show long term irregular
variability.  Second, there can be small magnitude shifts between seasons, and (for one field in
particular) the shifts were being misinterpreted as quasar-like variability. We carried out a quick visual
inspection of all candidates, but this was over kill and would certainly be automated in any 
subsequent analysis.  We were left with 1063 variability-selected candidates.

\begin{figure}
\centering
\includegraphics[width=8cm]{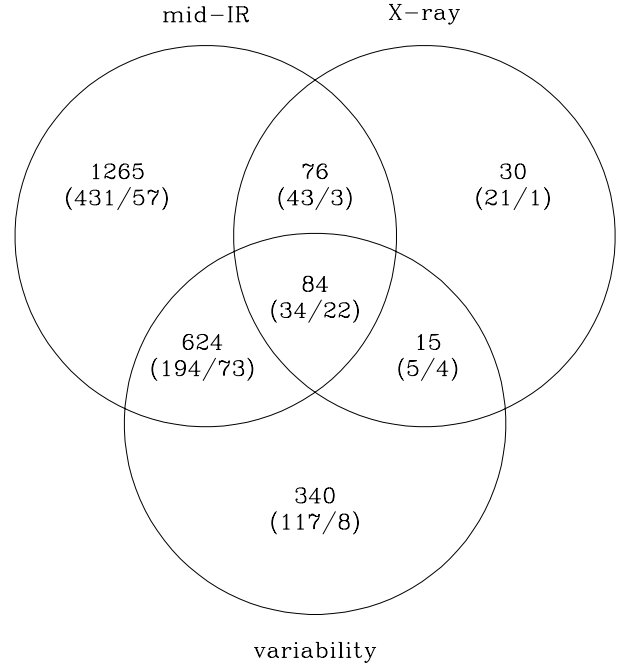}
\caption{Venn diagram for all 2434 LMC MQS quasar candidates. The upper numbers give
the number of candidates in each selection region. 
Priority 9 is given to AGN candidates located in the three overlapping circles (84 objects), 
priority 8 to objects in any two overlapping circles (715 objects), 
and priority 7 to sources in the non-overlapping areas (1635 objects). The lower
numbers, in parenthesis, give the number of objects in the four observed fields
followed by the number of spectroscopically confirmed AGN.}
\label{fig:VennCand}
\end{figure}

\subsection{Final AGN Candidates Sample}
\label{sec:finalsample}

The total number of candidates is 2434, distributed over $\sim$30 deg$^2$ in the LMC. There are 2049 mid-IR-selected 
candidates and 385 non-mid-IR-selected candidates.
Of the 2049 mid-IR (385 non-mid-IR) objects, 708 (355) are also variable and 160 (45) have associated X-ray emission. 
There are 1063 objects selected as variable, of which 708 are mid-IR sources, 99 are X-ray sources, and 84 both mid-IR and X-ray sources. 
Figure~\ref{fig:VennCand} shows how the samples overlap. Note that the limited overlap of the mid-IR and X-ray-selected AGNs is also seen 
in the spectroscopically confirmed samples of AGNs in the NDWFS field (e.g., \citealt{2011ApJ...728...56A}).
The AAOmega fiber allocation software allows to prioritize targets from 1 (the lowest priority) to 9. We gave priorities
of 9, 8, and 7 to sources selected by all three methods (84 sources), sources selected by only two methods (715 sources),
and sources selected by only one method (1635), respectively.

\section{Data Preparation and Observations}
\label{sec:data}

\begin{deluxetable}{lccccr}
\tabletypesize{\tiny}
\tablecaption{The MQS LMC Fields\label{tab:MQSfields}, Number of Candidates, Number of Confirmed Quasars and Exposure Times.}
\tablehead{Field & R.A. & Decl. & $N_{\rm cand}$ & $N_{\rm QSO}$ & $T_{\rm exp}$ (s)}
\startdata
LMC 01 & 04:41:43 & $-$69:50:29 & 220 & $\cdots$ & $\cdots$ \\
LMC 02 & 04:43:19 & $-$68:19:18 & 197 & $\cdots$ & $\cdots$ \\
LMC 03 & 04:56:52 & $-$67:07:48 & 213 & $\cdots$ & $\cdots$ \\
LMC 04 & 05:00:51 & $-$70:27:49 & 221 & 41 & 1480 \\
LMC 05 & 05:01:42 & $-$68:49:26 & 269 & $\cdots$ & $\cdots$ \\
LMC 06 & 05:14:27 & $-$67:35:00 & 189 & $\cdots$ & $\cdots$ \\
LMC 07 & 05:19:43 & $-$69:31:07 & 307 & 36 & 5400 \\
LMC 08 & 05:21:54 & $-$71:02:48 & 247 & 60 & 3600 \\
LMC 09 & 05:32:54 & $-$68:31:01 & 230 & $\cdots$ & $\cdots$ \\
LMC 10 & 05:41:31 & $-$71:36:00 & 201 & 37 & 2550 \\
LMC 11 & 05:41:56 & $-$70:11:07 & 275 & $\cdots$ & $\cdots$ \\
LMC 12 & 05:52:39 & $-$68:57:51 & 183 & $\cdots$ & $\cdots$ \\
\enddata
\tablecomments{Each field has a 1 deg radius. $N_{\rm QSO}$ does not have to add up to \newQSOs new QSOs and \knoQSOs known QSOs, 
since the fields overlap slightly (see Figure~\ref{fig:MQS_LMC_fields}) and one quasar can be observed in two or three fields.}
\end{deluxetable}

\begin{figure*}
\centering
\includegraphics[width=15cm]{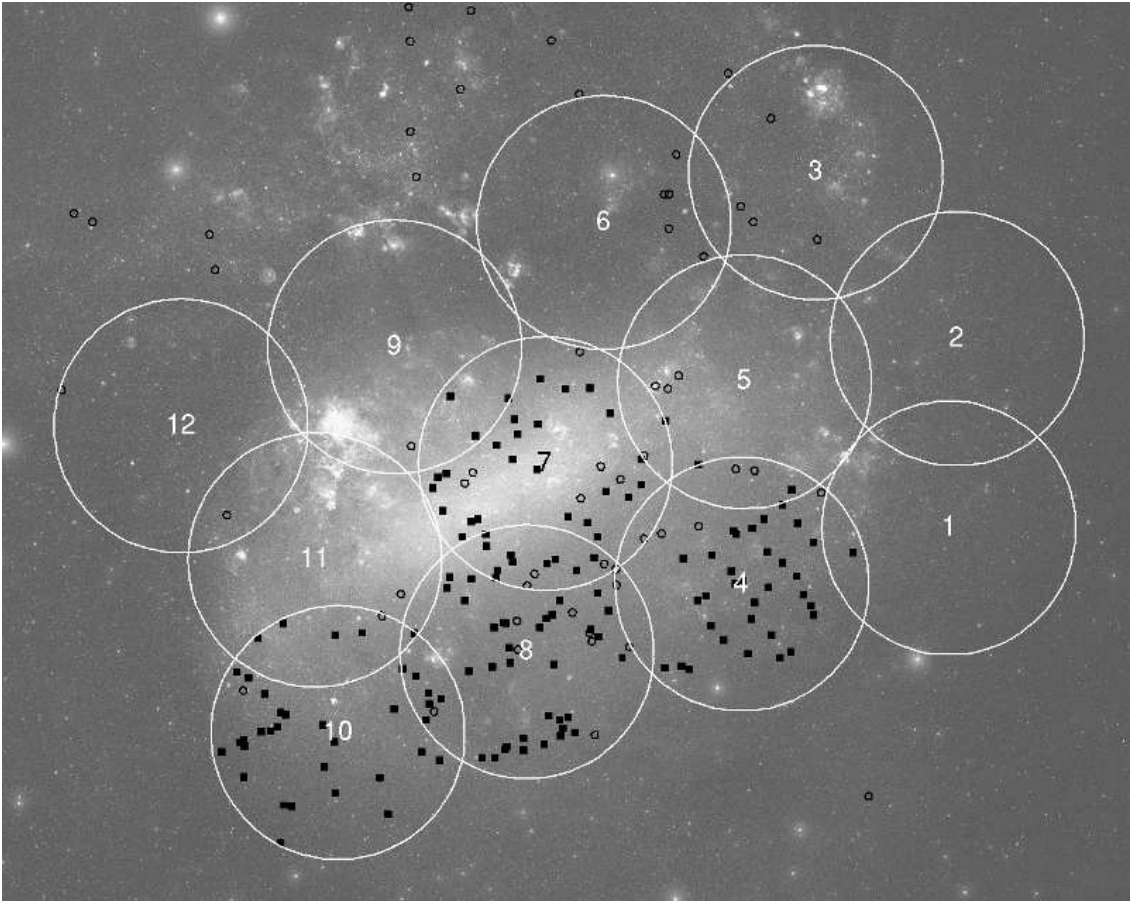}
\caption{Twelve MQS LMC fields (large circles). The small open circles mark 
the previously known quasars behind the LMC, while the filled squares mark the \newQSOs new quasars confirmed in this study.
Due to bad weather we observed only four (4, 7, 8, and 10) of 12 planned fields.
The image spans approximately 10 deg $\times$ 7 deg. North is up, east is to the left. }
\label{fig:MQS_LMC_fields}
\end{figure*}

We are mostly interested in quasars located in the overlapping areas of the SAGE and OGLE-III surveys. We divided this LMC area 
into 12 fields (see Figure~\ref{fig:MQS_LMC_fields}), each $\sim$3 deg$^2$ and corresponding to the field of view of AAOmega. 
The basic information for each field is given in Table~\ref{tab:MQSfields}.
The fields overlap slightly to avoid having gaps between them, so some of the quasar candidates were observed up to three times 
if they happened to fall into such an overlapping region. 
The total number of targeted objects in all 12 fields was 2678, of which several (less than 244) were targeted at least twice. 
On average there were 220 objects in a field, corresponding to $\sim$70 sources deg$^{-2}$.

We observed the LMC fields on 2011 February 1--3. 
Unfortunately, due to arrival of tropical storm Yasi, we were completely clouded out on the night of February 2 and partly on February 1 and 3.
Of the 12 planned fields, we were able to execute only four (see Table~\ref{tab:MQSfields} and Figure~\ref{fig:MQS_LMC_fields}), and only one with the desired 1.5 hr (5400 s) exposure time.
Nevertheless, the data are sufficient to confirm a large number of quasars. We observed \observed candidates of which we confirm \allQSOs (\newQSOs new) in this paper.
The yields divided by the AGN selection method are presented in Table~\ref{tab:selectionresults}.

We followed the spectral identification procedures from Paper I (\citealt{2011ApJS..194...22K}).
To find as many quasar emission lines in a spectrum as possible, we decided to use the widest available spectral coverage of $\sim$5100\AA\, (3750--8850\AA) and a resolution of $R=1300$.
Each spectrum consists of a blue (580V) and a red (385R) channel that were spliced together at 5700\AA. Some of the spectra appear to have a ``wide broad line'' 
around the splicing wavelength, mimicking the MgII (2800\AA) line at a redshift of $z\approx1.04$. We carefully checked all our $z\approx 1$ quasars for possible misidentifications.
Our initial goal was to obtain three exposures of $1800$~s per field, resulting in a median signal-to-noise ratio (S/N) of $\sim$25 at $I=19.5$ mag (calculated from obtained spectra). 
To subtract the sky flux contribution from each spectrum,
we initially selected $\sim$50 and finally used $\sim$25 sky fibers, each positioned to avoid stellar emission based on the OGLE-III LMC
catalogs (\citealt{2008AcA....58...89U}). These catalogs were also used to find $\sim$20 single and well separated bright guide stars for each field, of which 4--8 were used for guiding.
The AAOmega {\sc FLD} files were created with the {\sc Configure} software, and the data reduction was performed using the
{\sc 2dfdr} software (\citealt{1996ASPC..101..195T}).

As in Paper I, we searched each spectrum for the common (redshifted) quasar emission lines (e.g., \citealt{2001AJ....122..549V})
Ly$\alpha$ at 1216\AA, H$\delta$ at 4101\AA, H$\gamma$ at 4340\AA, H$\beta$ at 4861\AA, H$\alpha$ at 6563\AA, magnesium
MgII at 2800\AA, carbon CIV at 1549\AA~and CIII] at 1909\AA, as well as 
the narrow forbidden lines of oxygen [O II] at 3727\AA, [O III] at 4959\AA~or 5007\AA.
In general, we required the identification of two lines, except in the redshift range 
$0.7 < z < 1.2$, where we can only find MgII despite the broad spectral coverage.


\section{New Quasars}
\label{sec:newquasars}

We identified \newQSOs new quasars behind the LMC. Their basic parameters (including the identified lines) are presented
in Table~3. 
We also confirmed the \knoQSOs known quasars that were in our target list (Table~4).
Of these known quasars, we had flagged four only as probable quasars because their spectra were too
noisy for reliable classification.  We missed one known quasar, MQS~J051140.7$-$710032.8, that
had extremely low S/N in our data, although we verified its existence.
These five sources are reported at the bottom of Table~4.

\begin{figure}
\centering
\includegraphics[width=8cm]{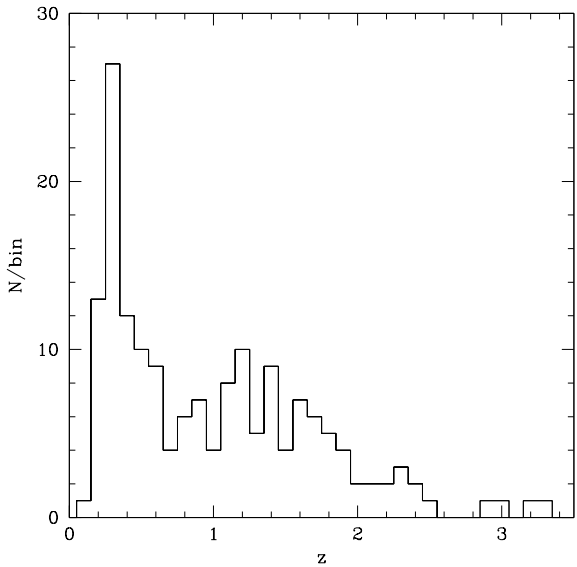}
\caption{Redshift distribution of the confirmed quasars in $z=0.1$ bins.}
\label{fig:hist}
\end{figure}

\begin{figure*}
\centering
\includegraphics[width=15cm]{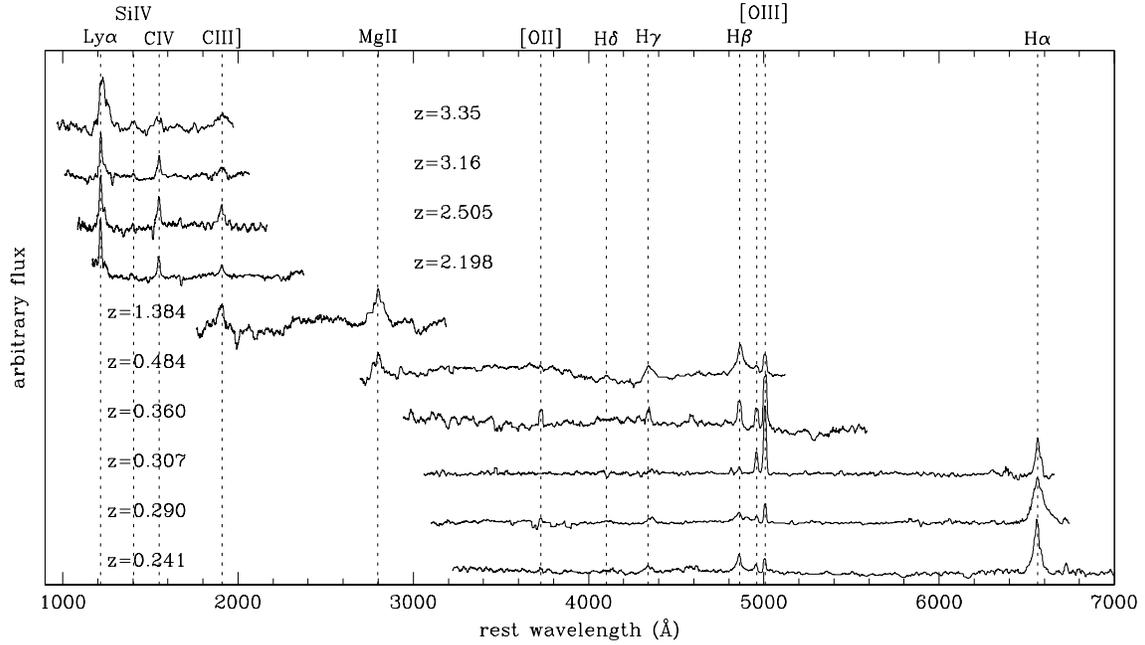}
\caption{Ten of the \newQSOs new LMC AGNs reported in this paper. 
The spectra have been flattened, smoothed, and scaled. 
The majority of the $z\approx0$ LMC emission lines as well as the atmospheric absorption features have been masked in order to emphasize the quasar emission lines. 
Each spectrum is labeled by redshift and we also mark the common quasar lines (vertical dashed lines with labels).}
\label{fig:spectra1}
\end{figure*}

\begin{figure*}
\centering
\includegraphics[width=15cm]{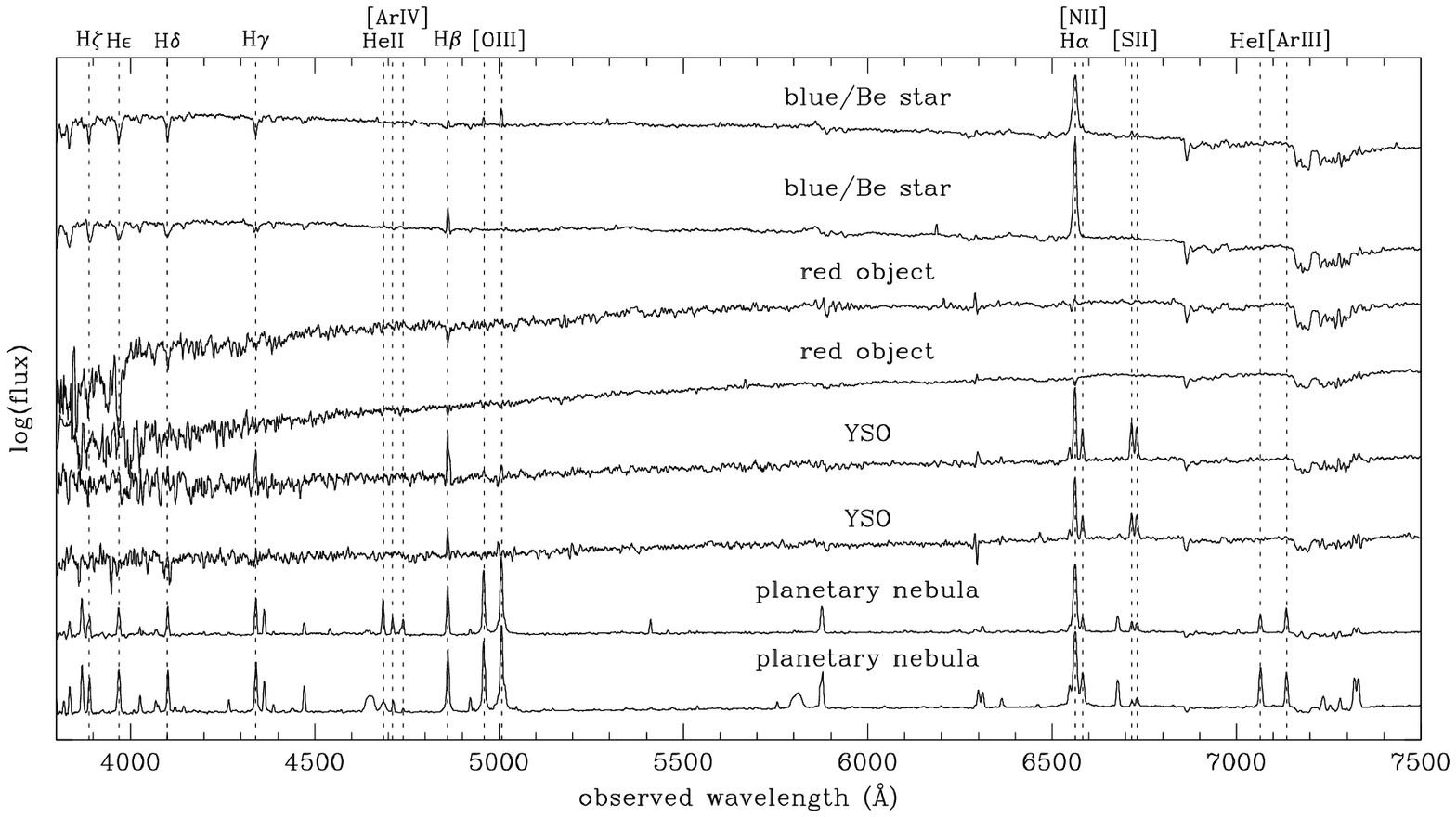}
\caption{Common contaminants can be divided into four groups: PNe, YSOs, red objects, and blue/Be stars. 
We show two spectra for each class, and also mark basic lines (vertical dashed lines with labels).}
\label{fig:contamination}
\end{figure*}

The newly discovered quasars cover the range of redshifts $z=0.15$--$3.35$ as shown in Figure~\ref{fig:hist}.
The deficit at $z\approx 0.8$ is probably due to problems in convincingly identifying quasars based on only MgII
in the region of the dichroic split. We show ten selected spectra of the
newly confirmed AGNs in Figure~\ref{fig:spectra1} and also show several spectra of
common contaminating sources in Figure~\ref{fig:contamination}. The remaining spectra are discussed in Section~\ref{sec:contamination}, and are
either not quasars or quasars with too low S/N for clear detection. 
As we discuss in Section~\ref{sec:discussion}, a fair number of the fainter sources ($I \gtrsim 19$ mag) should be quasars.

\begin{figure}
\centering
\includegraphics[width=8cm]{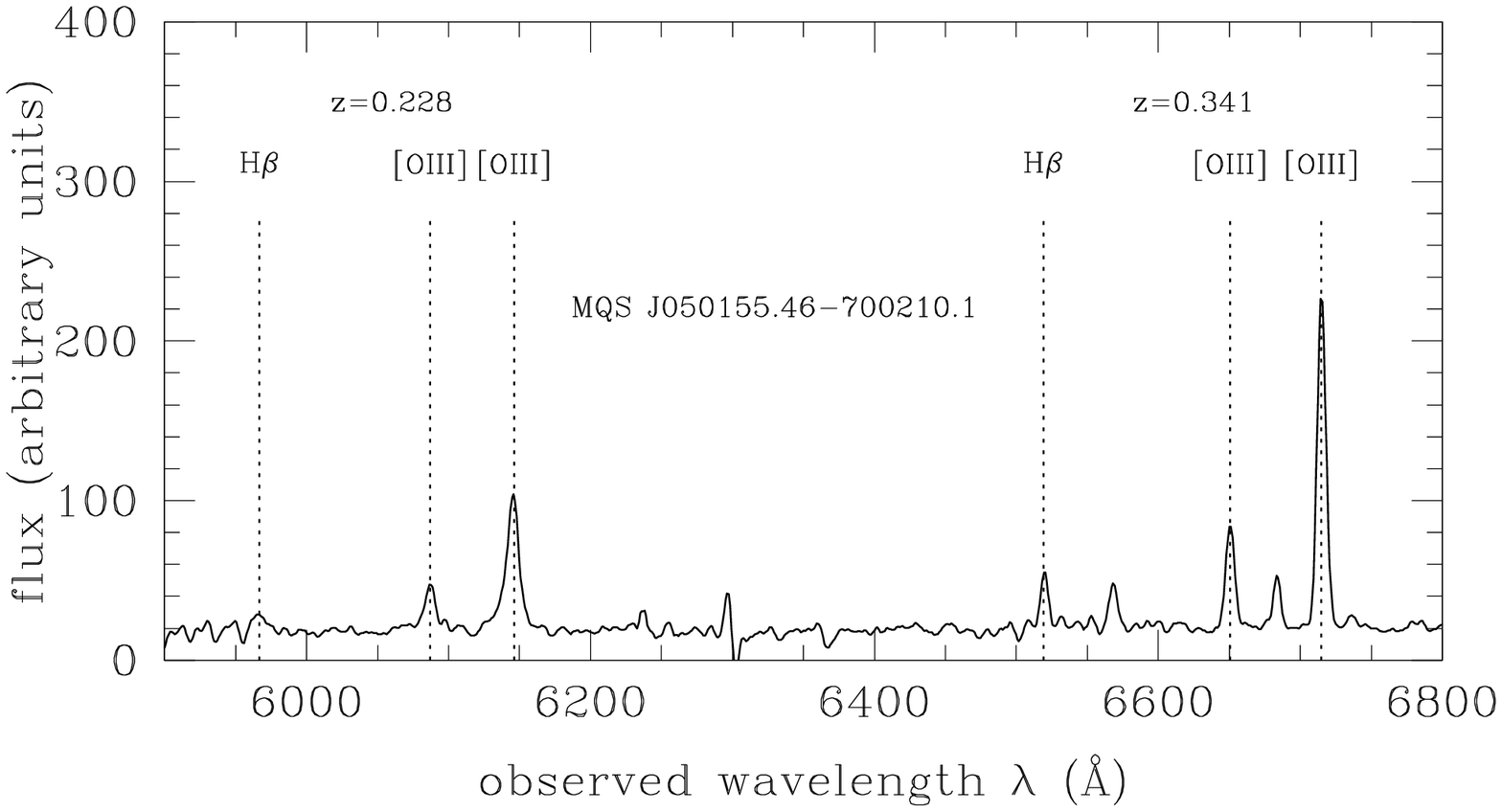}
\caption{Spectrum for MQS~J050155.46$-$700210.1. It displays two sets of 
the same emission lines at two redshifts, $z=0.228$ and $z=0.341$. 
We did not mask the LMC's ISM lines [OI] (6300\AA), H$\alpha$ 
(6563\AA), and HeI (6678\AA).}
\label{fig:doubleAGN}
\end{figure}

\begin{figure}
\centering
\includegraphics[width=8cm]{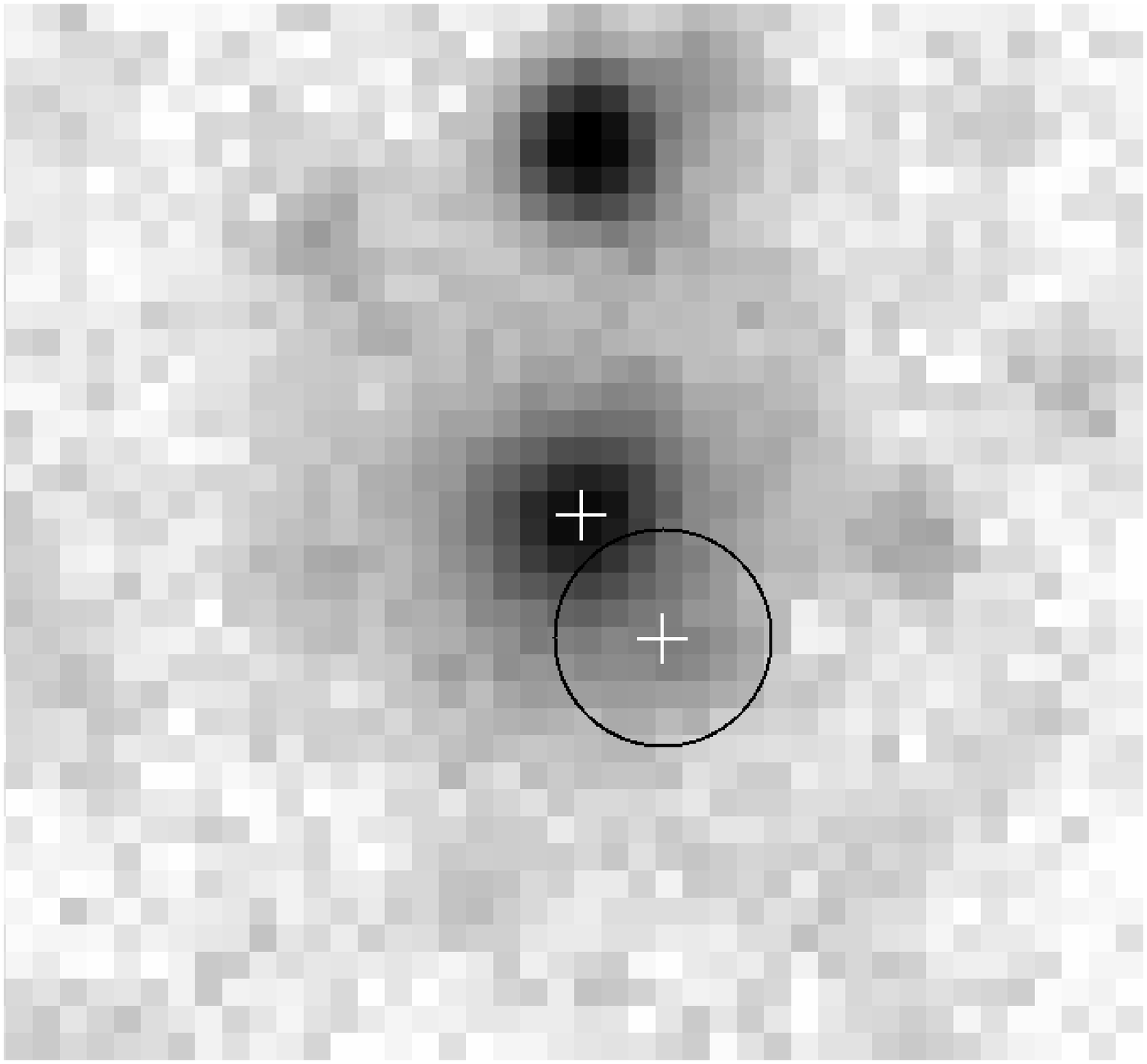}
\caption{Finding chart for MQS~J050155.46$-$700210.1. 
The circle has a diameter of 2 arcsec, corresponding to the AAOmega fiber size. 
The plus symbol inside the fiber marks the targeted source, 
while the plus symbol outside the fiber marks the galaxy, most likely hosting an AGN. 
The image is $10 \times 10$ arcsec. North is up, east is to the left.}
\label{fig:doubleAGNchart}
\end{figure}

One target, MQS~J050155.46$-$700210.1, shows emission lines corresponding to
redshifts of $z=0.228$ and $0.341$, as shown in Figure~\ref{fig:doubleAGN}.
This could be a chance coincidence
or a gravitational lens.  In the OGLE-III reference image
(Figure~\ref{fig:doubleAGNchart}), we find that the AGN candidate was a faint
source ($V=21.3$, $I=20.5$) in the wings of a brighter
extended source.  
The standard OGLE catalogs may not include all extended sources, 
so it is likely that the faint source was mistargeted
and the brighter galaxy is the AGN candidate. 
If we assume the bright galaxy is acting as a gravitational
lens, then we estimate that it has an Einstein radius of
$R_{\rm E}<0.4$ arcsec, which is very small because of the small redshift difference.
Higher resolution
images and spatially resolved spectra are required to clarify
the nature of this source.

\section{Contaminants}
\label{sec:contamination}

For the sources that were not quasars, we used our spectra of the known sources to
build template YSO, PN, blue/Be, and red star spectra, respectively,
and fit them to the individual spectra. 
Apart from the automatic fits, we also visually inspected each spectrum.
Table~5 reports the
object type for each of the sources and Figure~\ref{fig:contamination}
shows examples of some of these sources. 
We briefly describe each type below.

{\it Planetary nebulae.} We classified sources as candidate PNe if they had roughly equally
strong H$\alpha$, H$\beta$, and/or [OIII] emission lines along with several additional
lines such as HeI at 5876\AA, 6680\AA~and 7065\AA, HeII at 4686\AA, [ArIII] at 7136\AA, 
[NII] at 6548\AA~and 6584\AA, [SII] 6716\AA~and 6731\AA, and/or [ArIV] at 4711\AA~and 4740\AA~(see, e.g., \citealt{1991ApJS...75..407M}).
Of the 17 we classified as PNe based on our spectra, 7 were classified as such in SIMBAD and
5 were classified as YSOs.

{\it Young stellar objects.} We classified sources as candidate YSOs if they had significantly
weaker [OIII] and H$\beta$ lines relative to H$\alpha$ combined with relatively strong 
[SII] emission lines at 6716\AA~and 6731\AA, [OI] at 6300\AA, and/or 
[NII] at 6548\AA~and 6584\AA~(see, e.g., \citealt{1998AJ....115.2491K}). 
Of the 115 objects we classified as YSOs, 18 were classified as such by SIMBAD and 1 was classified as a PN,
leaving 96 new candidate YSOs.  There were also four additional sources we identify as
YSO/PN where we were uncertain as to the correct classification.  SIMBAD has two
of these four classified as YSOs.

{\it Be and blue stars}.  Blue stars had blue spectra with strong hydrogen Balmer series
absorption.  They were then further classed as Be stars if they had strong hydrogen
H$\alpha$ (H$\beta$) emission lines.  We identified 63 blue stars of which 39 appear
to be Be stars.  While there is little ambiguity about the Be stars, SIMBAD labels
two of the blue stars without emission lines as PNe, one as a Mira-type star
and one only as a Two Micron All Sky Survey (2MASS) infrared source (\citealt{2003yCat.2246....0C}).
We also found eight objects that could be either blue stars or YSOs.
The median OGLE-III color of these blue stars is $(V-I)\simeq0.1$ mag.

{\it Red stars}. These are stars with red spectra sometimes showing molecular absorption
bands.  We identified 68 red sources, where SIMBAD classifies 4 as infrared sources 
(one YSO from \citealt{2009ApJS..184..172G} and three 2mass sources), 
1 as a PN, 1 as a YSO, 1 as an asymptotic giant branch star, and 1 as a galaxy.
The median OGLE-III color of these sources is $(V-I)\simeq1.1$ mag.


\section{Discussion}
\label{sec:discussion}

\begin{figure}
\centering
\includegraphics[width=8cm]{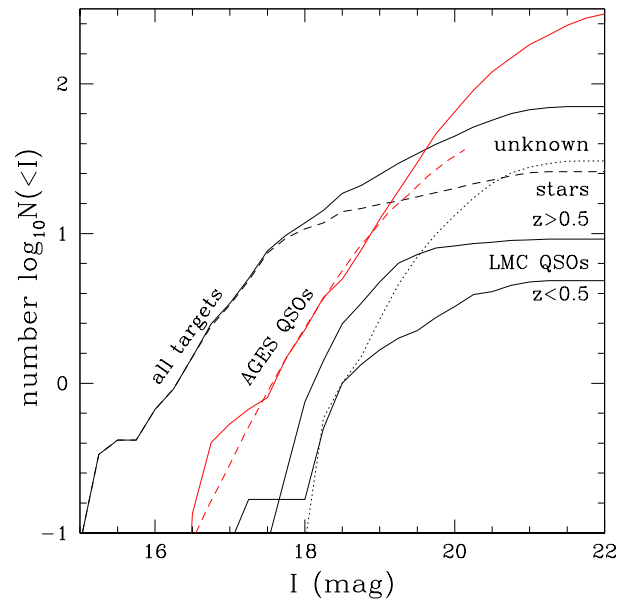}
\caption{Cumulative distributions of MQS targets.  The upper solid curve shows the distribution of all
the targets, which is then decomposed into stars (dashed), unclassified spectra (dotted), and quasars
(solid, divided into $z<0.5$ and $z>0.5$ sources).  The red solid (dashed) curves show the distribution
of $z>0.5$ AGES (SDSS $0.3 < z < 2.2$) quasars for comparison.  Our selection methods and bands are more
comparable to AGES than to SDSS.  }
\label{fig:cumul}
\end{figure}

The yield of our survey is determined by a combination of contamination and depth, 
where it is difficult to fully characterize the effects of depth because of the large 
variation in integration time created by the weather.
For discussion, we simply combine the four fields (Table~\ref{tab:selectionresults}).
We know from the surface density of candidates compared to quasars 
(\citealt{2011arXiv1110.4371K}, \citealt{2006AJ....131.2766R}), 
shown in Figure~\ref{fig:cumul}, that the level of contamination is high, 
but much of this is by design because of the large number of available fibers.

\begin{deluxetable}{lccc}
\tabletypesize{\tiny}
\tablecaption{MQS LMC Yields\label{tab:selectionresults}. }
\tablehead{Selection & Observed & Confirmed & Yield \\
 & Targets & AGNs & \%}
\startdata
Mid-IR QSO-Aa & 577 & 136 (137$^*$) & 24 \\
Mid-IR QSO-Ab & 10 & 2 & 20 \\
Mid-IR QSO-Ba & 62 & 12 & 19 \\
Mid-IR QSO-Bb & 1 & 0 & 0 \\
Mid-IR YSO-Aa & 32 & 4 & 13 \\
Mid-IR YSO-Ab & 16 & 1 & 6 \\
Mid-IR YSO-Ba & 1 & 0 & 0 \\
Mid-IR YSO-Bb & 3 & 0 & 0 \\
\hline
Mid-IR (any) & 702 & 155 (156$^*$) & 22 \\
X-ray (any) & 102 & 30 & 29 \\
Var. (any) & 350 & 107 & 31 \\
X-ray + Mid-IR (any) & 77 & 25 & 32 \\
Mid-IR + Var. (any) & 228 & 95 & 42 \\
Var. + X-ray (any) & 39 & 26 & 67 \\
\hline
\\
\multicolumn{4}{c}{Priority 7}\\
\hline
Mid-IR (only) & 431 & 57 (58$^*$) & 13 \\
X-ray (only) & 21 & 1 & 5 \\
Var. (only) & 117 & 8 & 7 \\
\hline
\\
\multicolumn{4}{c}{Priority 8}\\
\hline
X-ray + Mid-IR (only) & 43 & 3 & 7 \\
Mid-IR + Var. (only) & 194 & 73 & 38 \\
Var. + X-ray (only) & 5 & 4 & 80 \\
\hline
\\
\multicolumn{4}{c}{Priority 9}\\
\hline
All three & 34 & 22 & 65
\enddata
\tablecomments{$^*$MQS~J050155.46$-$700210.1 is consistent with having two redshifts (see, \S\ref{sec:newquasars}).}
\end{deluxetable}

In Figure~\ref{fig:VennCand} and Table~\ref{tab:selectionresults}, we present our yields from this observing run, divided by the selection method.
The mid-IR-selected sources were divided into several classes (QSO-Aa, QSO-Ab, etc.; see Section~\ref{sec:mIRsel}). 
As expected the purest class, QSO-Aa, has the highest confirmation rate, at 24\%. 
Next, the two classes QSO-Ab and QSO-Ba had confirmation rates of 20\% and 19\%, respectively.
The YSO-Aa and YSO-Ab groups have lower rates of 13\% and 6\%, as expected
from the original division into YSO and QSO classes.
Variability selection had a yield of 31\% and X-ray selection 30\%.

Priority 7 objects, where the source was flagged based on only one of the mid-IR, variability or X-ray criteria,
had confirmation rates of 13\% (57 new AGNs), 7\% (8), and 5\% (1), respectively. 
Of the priority 8 objects selected by any two methods but not the third, the combination of
X-ray plus variability yielded the highest confirmation rate, at 80\% (4 new AGNs), most likely due to low number statistics. 
The mid-IR and variability criteria had a yield of 38\% (73 AGNs) and X-ray plus mid-IR criteria had a yield of only 7\%, 
again with low number statistics.
Not surprisingly, combining all three selection methods was the most efficient with a yield of 65\% (22 new AGNs).

\begin{figure}
\centering
\includegraphics[width=8cm]{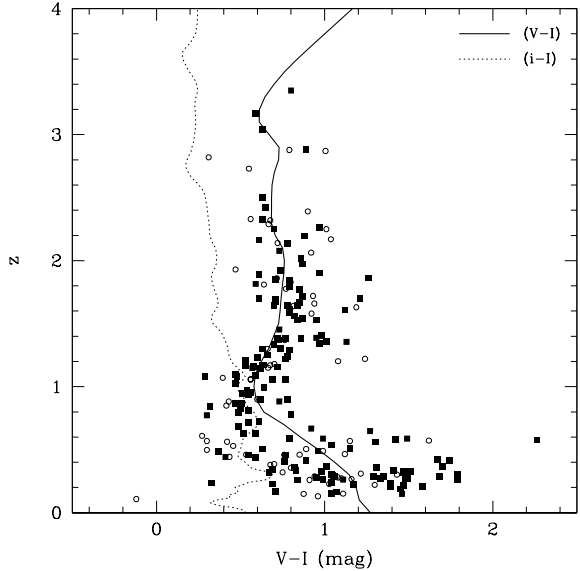}
\caption{$V-I$ colors of the MQS LMC quasars (filled squares) and other confirmed LMC and SMC quasars (open circles) as a function of redshift. 
Overplotted is the color for a pure AGN spectrum from \cite{2001AJ....122..549V} convolved with the $V$ and $I$ filters (solid line). 
The same spectrum was also convolved with the $i$ and $I$ filters in order to convert OGLE magnitudes into SDSS $i$-band magnitudes---the $(i-I)$ color is shown as dotted line.}
\label{fig:AGNcolor}
\end{figure}

Figure~\ref{fig:cumul} shows the cumulative distribution of our targets and then their division into quasars,
stars, and unclassified objects.  We also show the cumulative distribution of $z>0.5$ quasars in 
AGES (\citealt{2011arXiv1110.4371K}).  These quasars are primarily mid-IR and X-ray selected,
like much of the present sample, but with a much deeper optical spectroscopic limit.  We have adjusted
the AGES magnitudes for the $\Delta E(B-V)\simeq 0.14$~mag extinction difference between AGES and a typical LMC 
site line (\citealt{1999AcA....49..201U}), although this may underestimate the correction necessary for
a true background population.  The completeness of our present sample relative to AGES is roughly 
50\% (30\%) at 18.6~mag (19.3~mag). 
Figure~\ref{fig:cumul} also shows the cumulative distribution of the SDSS
quasars with $0.3 < z < 2.2$ from \cite{2006AJ....131.2766R}.  Given the change in bandpass (the
$(i-I) \simeq 0.5$~mag color term varies significantly with redshift, see Figure~\ref{fig:AGNcolor}) 
and the additional line width and luminosity selection criteria in \cite{2006AJ....131.2766R} it is harder 
to make a direct comparison, but after matching as best possible, we obtain similar completeness estimates.
As expected from filling the fibers with targets, most targets are peculiar stars of various types, 
particularly at the brighter magnitudes.  Unclassifiable spectra dominate at the fainter magnitudes,
and with the weather-limited integration times, there was an effective magnitude
limit of roughly $19.5$--$20$~mag.

Figure~\ref{fig:AGNcolor} shows the colors of the confirmed quasars as a function of redshift.
While we imposed no optical color selection criteria, the quasars generally follow the colors of SDSS quasars
except at very low $z$, where the host dominates the colors. Adding an optical color constraint 
has no leverage on the selection of quasars, as the colors of quasars are indistinguishable from the colors of many LMC stars.

\begin{figure*}
\centering
\includegraphics[width=15cm]{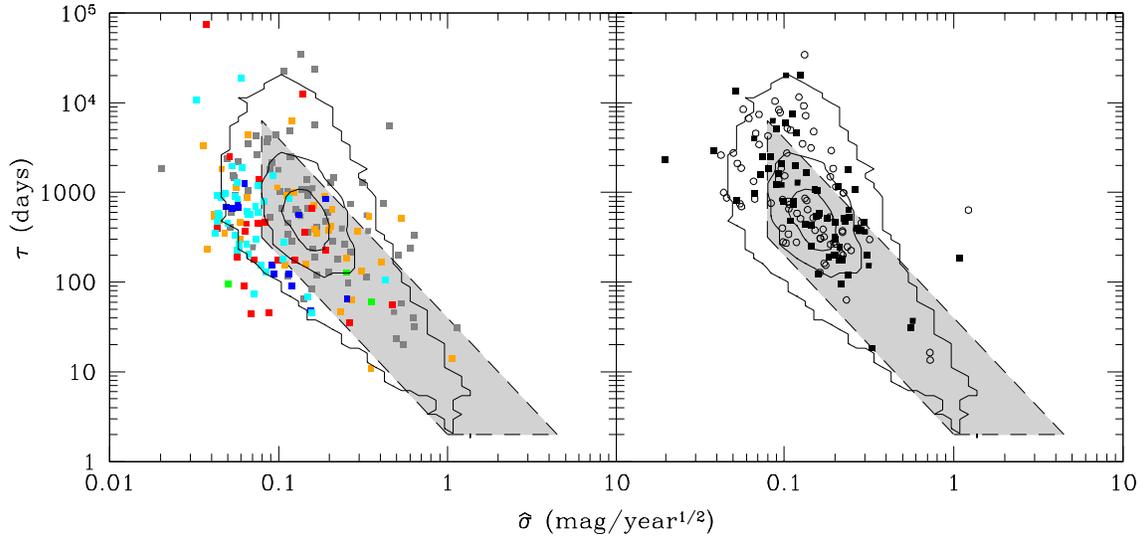}
\caption{$\tau$--$\hat{\sigma}$ (timescale--scaled amplitude) variability plane. 
In the left panel, we show the observed variable AGN candidates that are not quasars.
They are  YSOs (orange), blue stars (blue), red stars (red), PNe (green), Be stars (cyan), and objects with featureless spectra (dark gray).
In the right panel, we show the confirmed MQS LMC quasars (filled squares) and other known 
LMC and SMC quasars (open circles) as compared to the selection criteria in \cite{2010ApJ...708..927K} (gray area) 
and density contours (1, 10, and 20 per 0.1 dex bins in both axes) 
for $\sim$9000 variable SDSS AGNs from \cite{2010ApJ...721.1014M}.
The \cite{2010ApJ...708..927K} cut was designed to return high purity samples 
given the variability properties of contaminating stars. 
With plenty of fibers, however, we significantly extended this selection region (see Section~\ref{sec:varsel}).}
\label{fig:tau-sigma}
\end{figure*}

\begin{figure*}
\centering
\includegraphics[width=15cm]{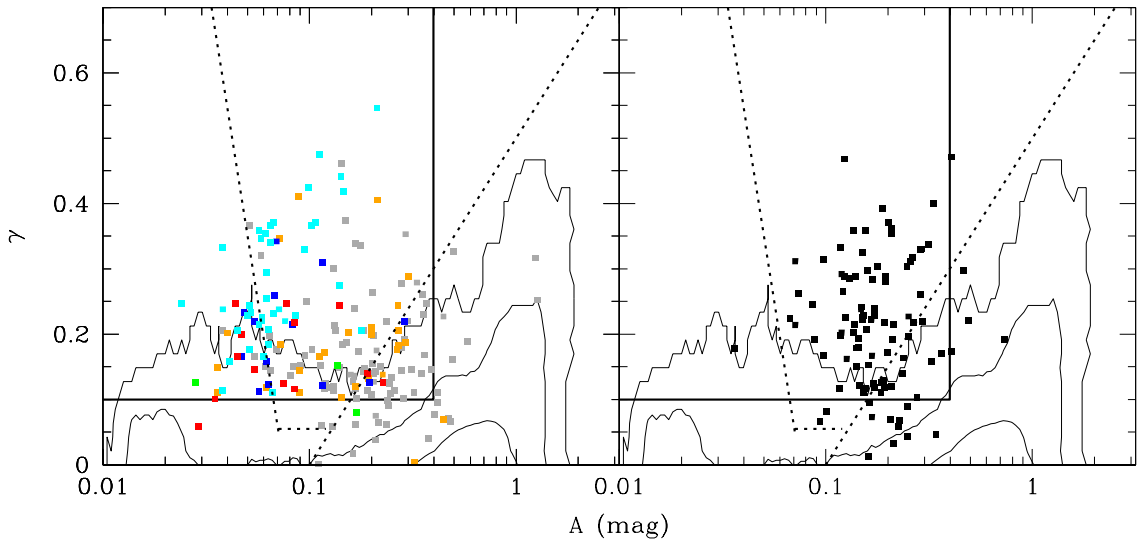}
\caption{$A$--$\gamma$ (amplitude--structure function slope) variability plane from \cite{2010ApJ...714.1194S}. 
In the left panel, we show the observed variable AGN candidates that are not quasars.
Color coding is as in Figure~\ref{fig:tau-sigma}.
In the right panel, we show the confirmed MQS LMC quasars (filled squares).
Our selection region is above the horizontal ($\gamma>0.1$) and left of the vertical ($A<0.4$ mag) solid lines. 
Some of our new confirmed quasars and quasar candidates are outside this cut---these are objects selected by methods other than variability.  
The dashed lines are the selection cuts from \cite{2010ApJ...714.1194S}.
The contours are for $\sim$30,000 variable objects from a 1 deg$^2$ area with the typical stellar density of the LMC. 
The objects are counted in $\Delta A=0.02$ dex and $\Delta \gamma=0.02$ bins. The outer, middle, and inner contours are for 1, 10, and 100 objects per bin, respectively.}
\label{fig:Schmidt}
\end{figure*}

\begin{figure*}
\centering
\vspace{0.3cm}
\includegraphics[width=15cm]{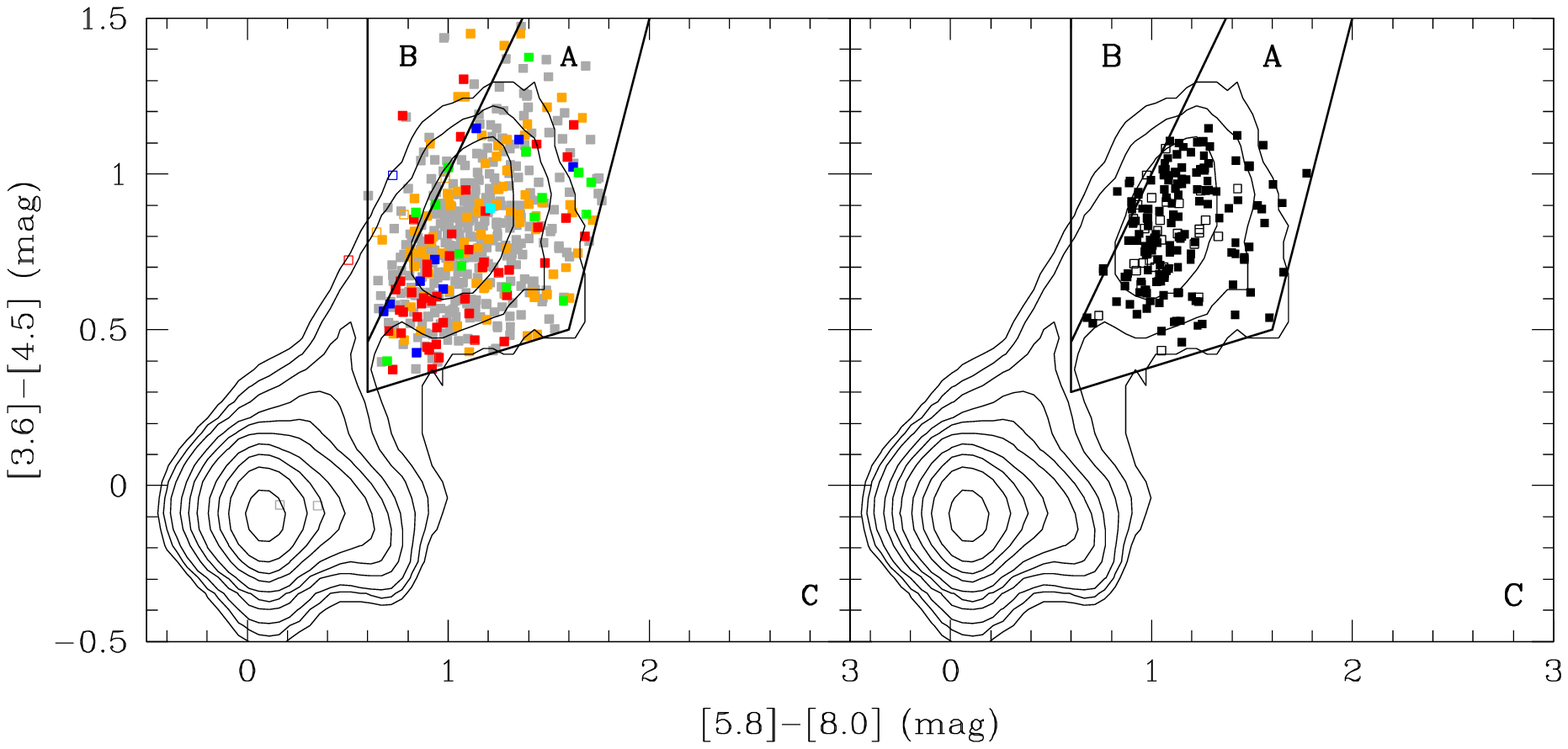}\\
\vspace{0.3cm}
\includegraphics[width=15cm]{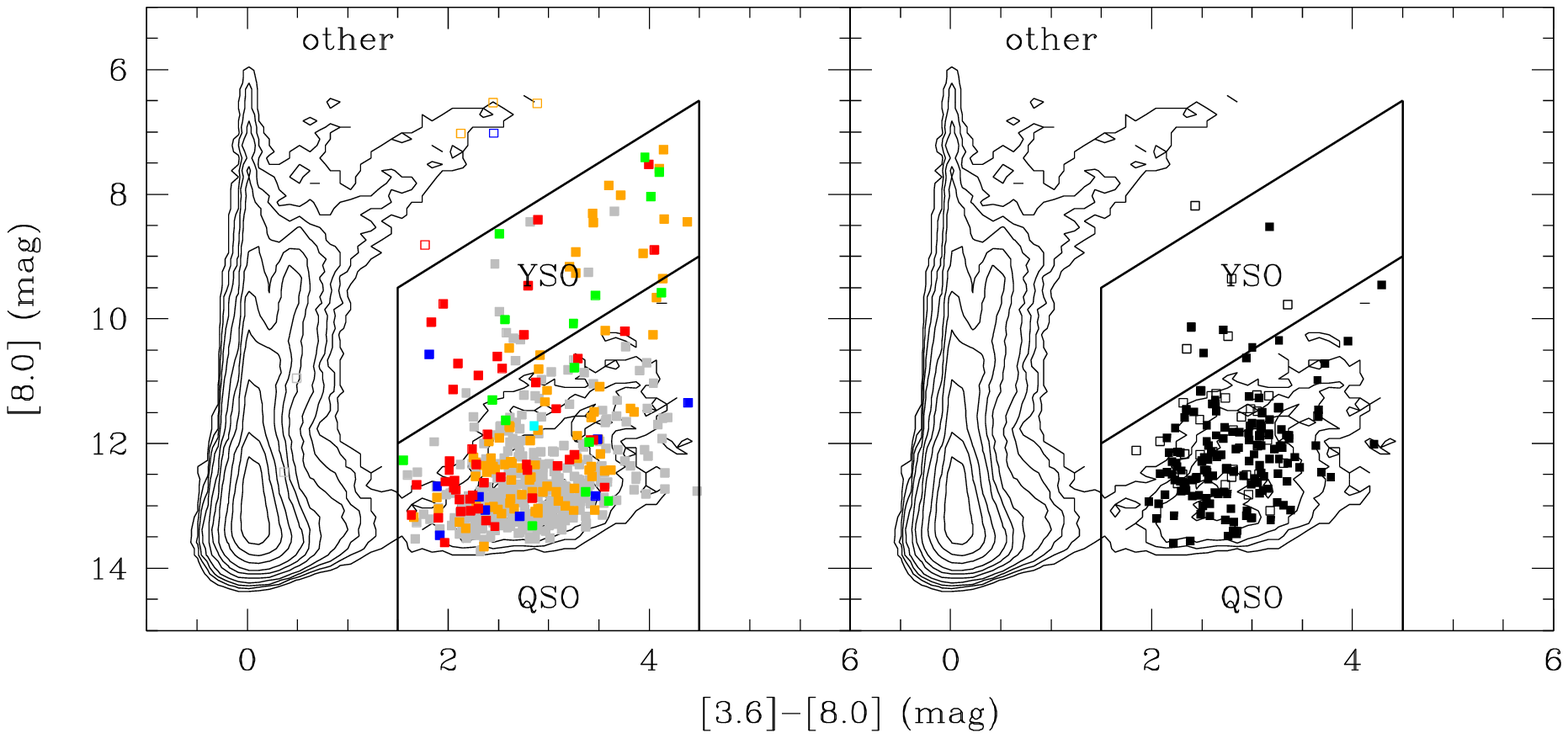}\\
\vspace{0.3cm}
\includegraphics[width=15cm]{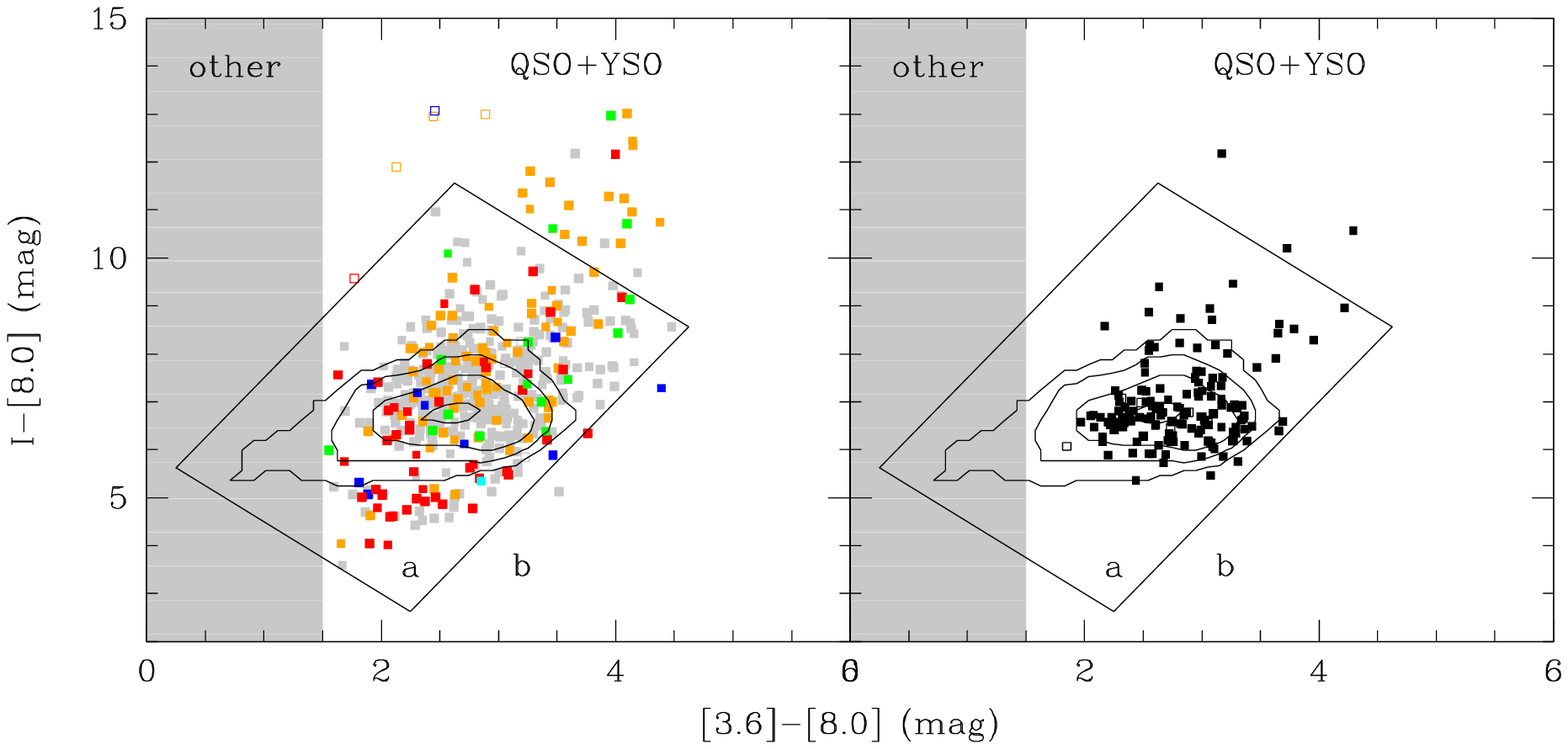}
\caption{AGN selection criteria from \cite{2009ApJ...701..508K}. These cuts are summarized in Section~\ref{sec:mIRsel}.
In the left panels, the filled symbols mark the mid-IR-selected confirmed YSOs (orange), blue stars (blue), red stars (red), 
PNe (green), Be star (cyan) and objects with featureless spectra (gray), while the open symbols are 
for the candidates selected by the variability and/or X-ray selection method and then matched to the {\it Spitzer} and OGLE-III data.
In the right panels, the filled symbols mark the confirmed quasars from this work (both new and known) and the open symbols 
are for the other known LMC quasars, that are outside of the four observed fields.}
\label{fig:KKO9}
\end{figure*}

In Figures~\ref{fig:tau-sigma}--\ref{fig:KKO9}, we examine the selection spaces used to select the candidates.
First, we inspect the $\tau$--$\hat{\sigma}$ variability plane from \cite{2010ApJ...708..927K} and \cite{2010ApJ...721.1014M}.
\cite{2010ApJ...708..927K} proposed a narrow and long trapezoidal region to select high purity sample of AGN candidates. 
Indeed, if we had restricted our variability selection to this region, we would have a yield of 70\% at the price
of a significant reduction in completeness.  In fact, the distribution of the variability parameters of all quasars
follows the distribution found by \cite{2010ApJ...721.1014M} for SDSS quasars, extending outside the \cite{2010ApJ...708..927K}
selection boundaries, particularly for long $\tau$ where there is also increased contamination from irregularly
varying stars.  The left panel of Figure~\ref{fig:tau-sigma} shows the distribution of objects classified as non-quasars (PNe, YSOs, etc.)
or where we failed to classify the spectrum in this variability plane. We note that the vast majority of classified non-quasars
lie outside the \cite{2010ApJ...708..927K} region.

In Figure~\ref{fig:Schmidt}, we make the same comparison between the MQS quasars and the other candidates 
in the $A$--$\gamma$ variability plane of \cite{2010ApJ...714.1194S}. 
We used a cut in this plane to remove more variable sources (shown as contours) other than quasars that we included as we
greatly relaxed the DRW selection criteria of \cite{2010ApJ...708..927K} in these dense stellar fields.
A fair number of common variable stars are periodic with the index of the power-law SF of $\gamma < 0.1$.
On the other hand, objects with very high amplitude variations 
are unlikely to be quasars as quasar variability amplitudes on long time-scales are expected to be $A\approx 0.3$ mag (e.g., \citealt{2010ApJ...721.1014M}).
Our simple cut (solid lines) allows higher contamination than that of \citet[dotted lines]{2010ApJ...714.1194S}, but also returns the higher number of confirmed quasars.

We also examined the distribution of the confirmed MQS quasars and quasar candidates  
relative to the mid-IR selection criteria of \citet[Figure~\ref{fig:KKO9}]{2009ApJ...701..508K}.
The quasars lie in the proposed selection regions confirming the basic principles outlined in \cite{2009ApJ...701..508K}.
The remaining quasar candidates (shown in the left panels) are either quasars with low S/N or contaminating sources. 
In the mid-IR color-color plot (top panels in Figure~\ref{fig:KKO9}), 
the confirmed quasars (shown in the right panel) and the faint candidates (gray squares in the left panel) occupy
nearly identical regions, while the blue and red stars tend to clump in the lower-left region of the ``AGN wedge'' area. 
Also, YSOs tend to have a somewhat larger color spread than the more centrally clumped AGNs. 
In the mid-IR CMD (middle panels of Figure~\ref{fig:KKO9}),
the confirmed quasars seem to be on average brighter than the remaining faint sources. 
The red stars tend to have bluer mid-IR $[3.6]-[8.0]$ colors than the other sources.
In the bottom panels of Figure~\ref{fig:KKO9}, we show the optical-to-mid-IR colors of our observed targets.
As expected the vast majority of our quasars occupy the ``a'' (QSO-like optical-to-mid-IR colors) region, but then
so do the most of other sources.  The red stars again have bluer optical-to-mid-IR colors than the other sources.

As expected from earlier surveys of mid-IR-selected quasars 
(\citealt{2005ApJ...631..163S,2010ApJ...713..970A}),
we find that 88\% of the mid-IR-selected quasars show optical
broad lines and are not Type~II quasars.  Similarly, of the
155 confirmed, mid-IR-selected quasars, 95 (61\%) are clearly
variable sources.


\section{Summary}
\label{sec:summary}

We observed \observed AGN candidates in four $\sim$3 deg$^2$ LMC fields with AAOmega to confirm 
\allQSOs quasars, including \newQSOs new ones. 
Given the 55 previously known quasars behind the LMC, including \knoQSOs in our present fields,
we have quadrupled (to 200) the number of known quasars behind the LMC.
By observing the eight remaining fields, we should find another $\sim$300 new quasars,
although the yield would increase if all data were obtained with the full integration time.

We observed the quasar candidates based on the priorities set by the number of the selection 
methods that picked them. 
If they were selected by only one method, either mid-IR colors, X-ray emission, 
or photometric variability, the average confirmation yield is low, at $\sim$9\%.
If the candidate was selected based on two methods but not the third, the average confirmation rate is at 40\%.
The highest yields come with the simultaneous selection by all three methods, with the yield at 65\%.
Note, however, that we deliberately chose to fill all the AAOmega fibers with candidates 
in order to achieve higher completeness, guaranteeing high contamination rates because 
the surface density of quasars is well below the surface density of fibers.

Our biggest problem given the availability of fibers is not contamination.  
We could assign rough classifications to
many contaminating sources, including 115 YSOs (96 new), 17 PNe (5 new), 
39 new Be and 24 blue stars, 68 red stars, and also
12 objects that are either YSO/PN or blue star/YSO.
The big problem is that we could not classify
402 spectra due to low S/N, what
is also a major problem for making black hole mass estimates
from the line widths of the quasar emission lines.

Nonetheless, tripling the number of quasars in the LMC and SMC
using only 5 hr on sky is a strong vindication of the general
approach and the advantages created by the large numbers of AAOmega fibers.  
With no improvements in yields due to actually obtaining
our target integration times, we estimate that completing the OGLE-III 
LMC and SMC fields would yield $\sim$700 quasars and completing the OGLE-IV
fields would yield $\sim$3600 quasars. 
These remain the best
fields for extending studies of quasar variability independent of
their additional uses for understanding the Galaxy through the proper 
motions of the LMC or probing the local ISM.


\acknowledgments

We thank Sarah Brough, our support astronomer, for help during the observing run. 
We thank Micha{\l} Jaroszy{\'n}ski for the helpful discussion on MQS~J050155.46$-$700210.1.
We also thank the anonymous referee, whose comments helped us to improve the manuscript.
This research was made based on observations with the Anglo-Australian Telescope, 
for which the observing time was granted by the Optical Infrared Coordination Network for Astronomy (OPTICON).
This research has made use of the SIMBAD database, operated at CDS, Strasbourg, France. 
This research has also made use of the NASA/IPAC Extragalactic Database (NED) which is operated by the Jet Propulsion Laboratory (JPL), 
California Institute of Technology (Caltech), under contract with the National Aeronautics and Space Administration (NASA). 

Support for this work was provided by the Polish Ministry of Science and Higher Education (MNiSW)
through the program ``Iuventus Plus'', award number IP2010 020470 to S.K., A.M.J., and A.U.
This work has been supported by NSF grants AST-0708082 and AST-1009756 to C.S.K.
The OGLE is supported by the European Research Council under 
the European Community's Seventh Framework Programme (FP7/2007-2013),
ERC grant agreement no. 246678 to A.U.


\clearpage

\setcounter{table}{2}

\begin{figure*}
\centering
\includegraphics[]{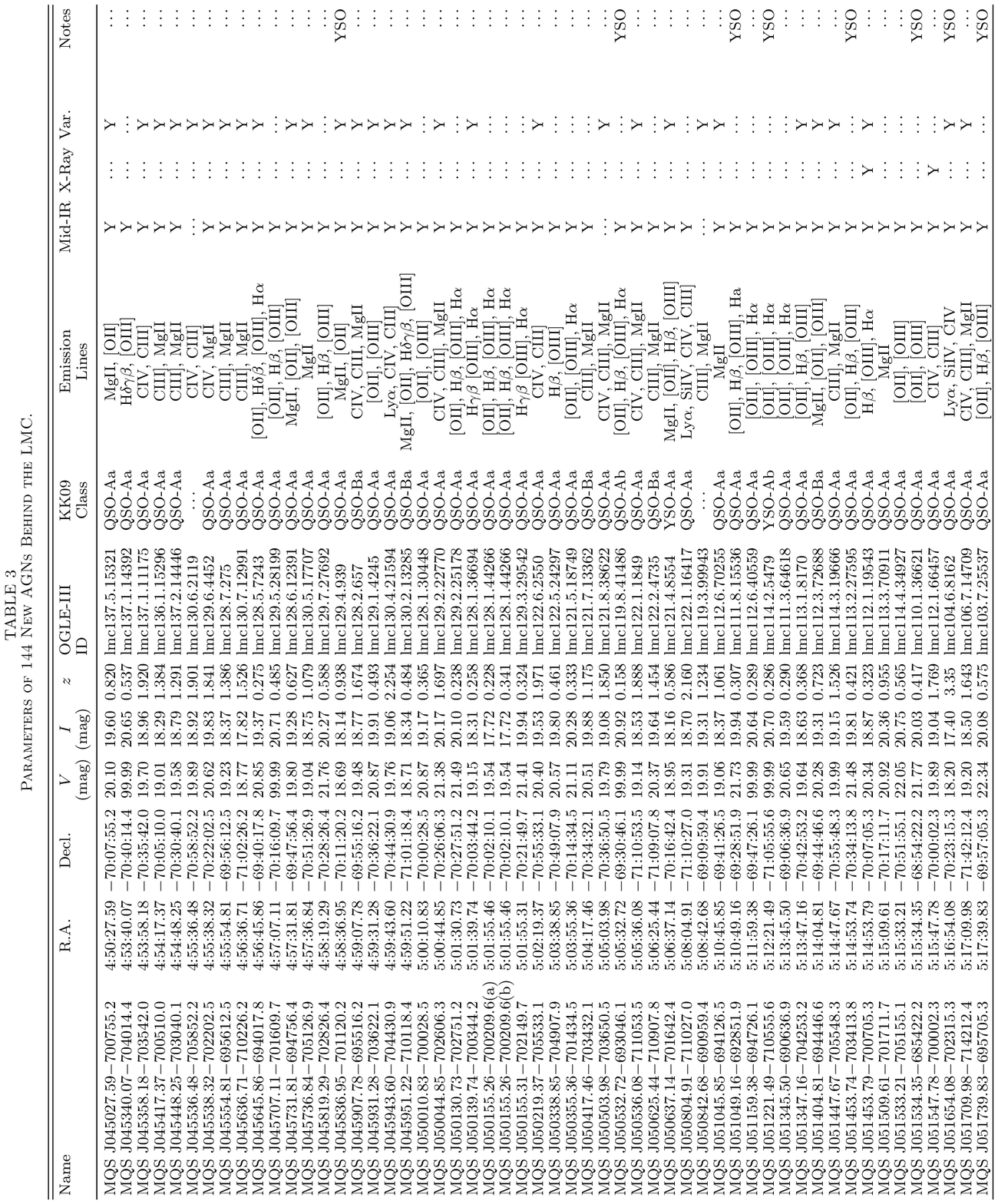}
\end{figure*}

\begin{figure*}
\centering
\includegraphics[]{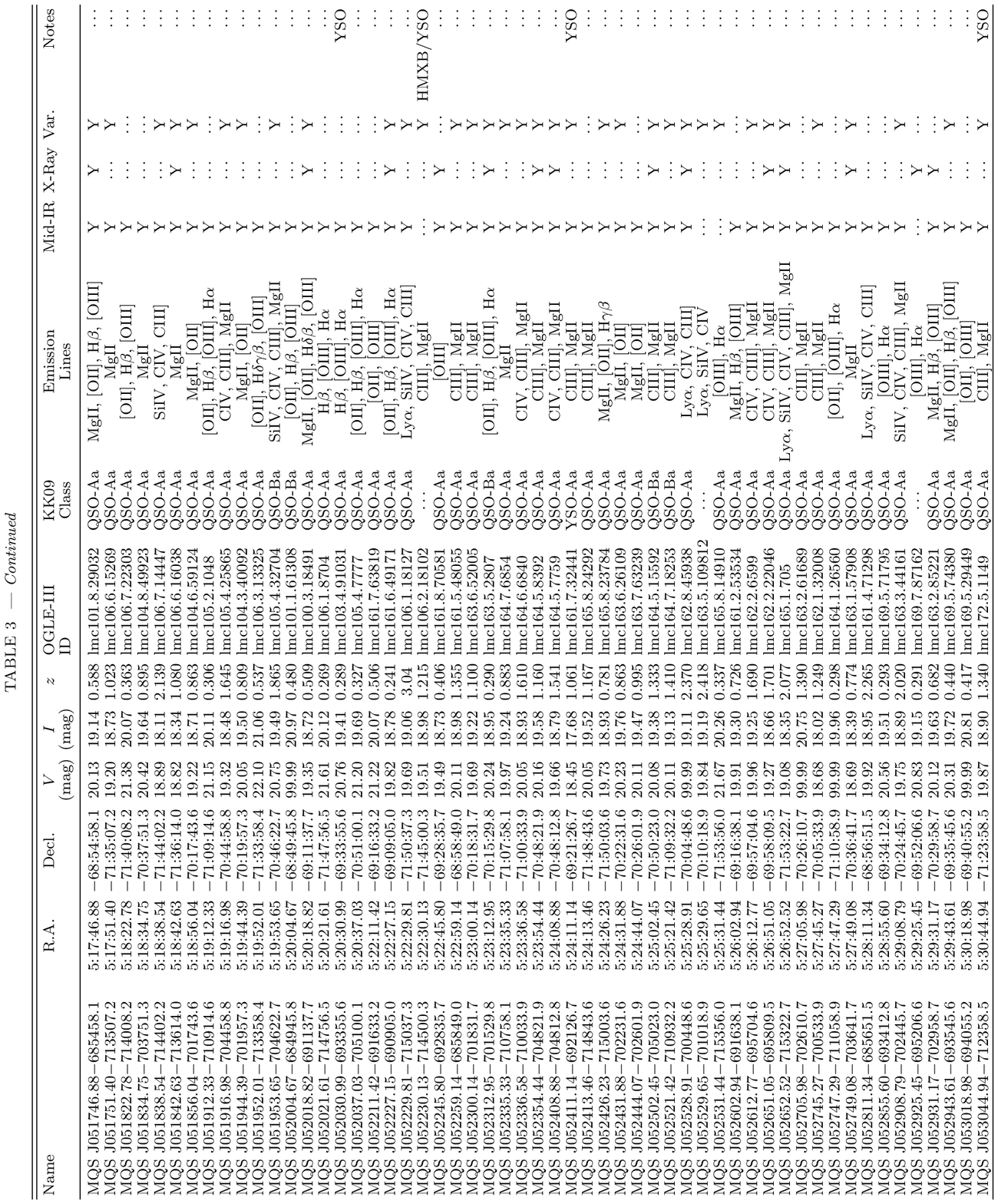}
\end{figure*}

\begin{figure*}
\centering
\includegraphics[]{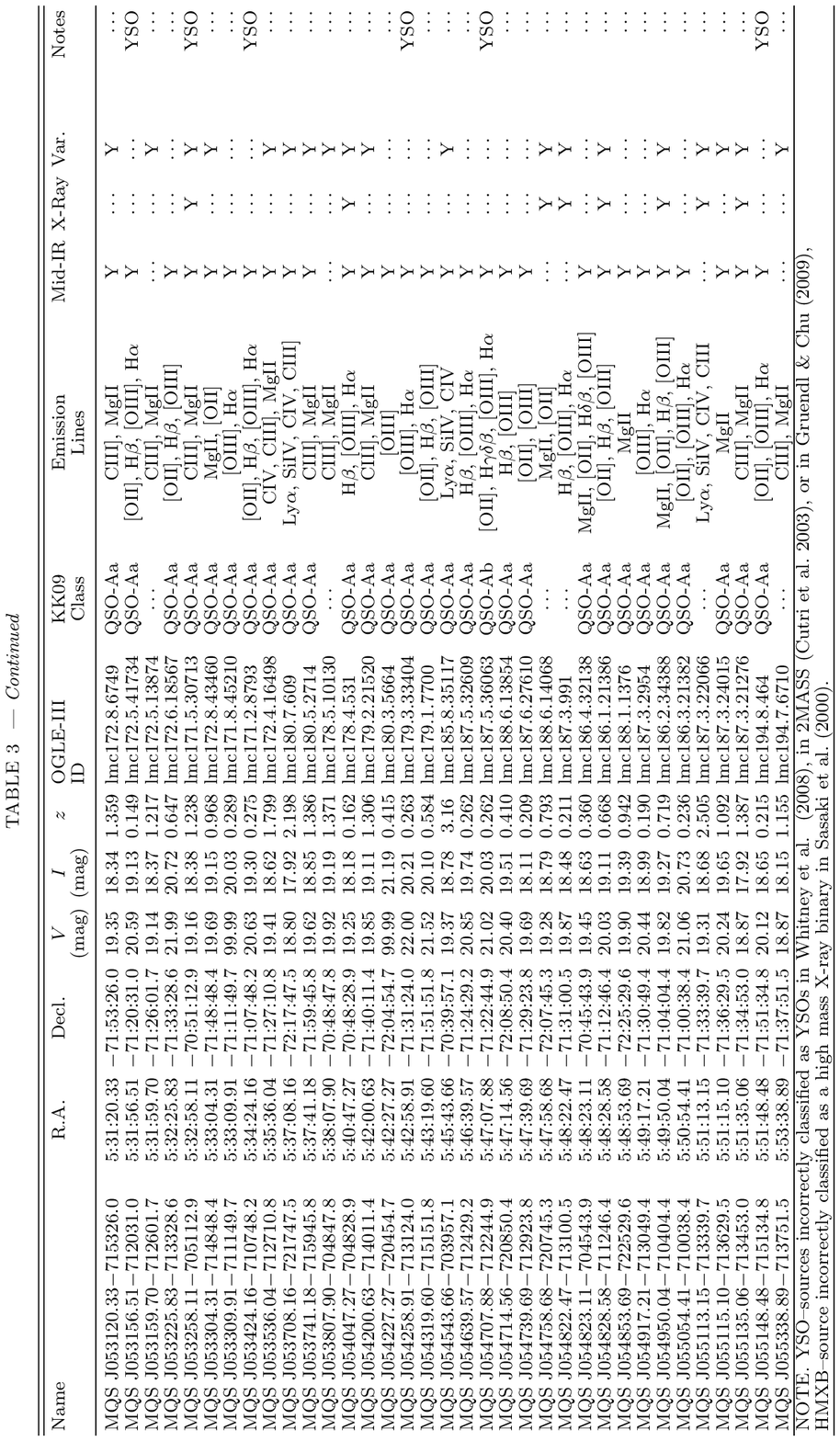}
\end{figure*}

\begin{figure*}
\centering
\includegraphics[]{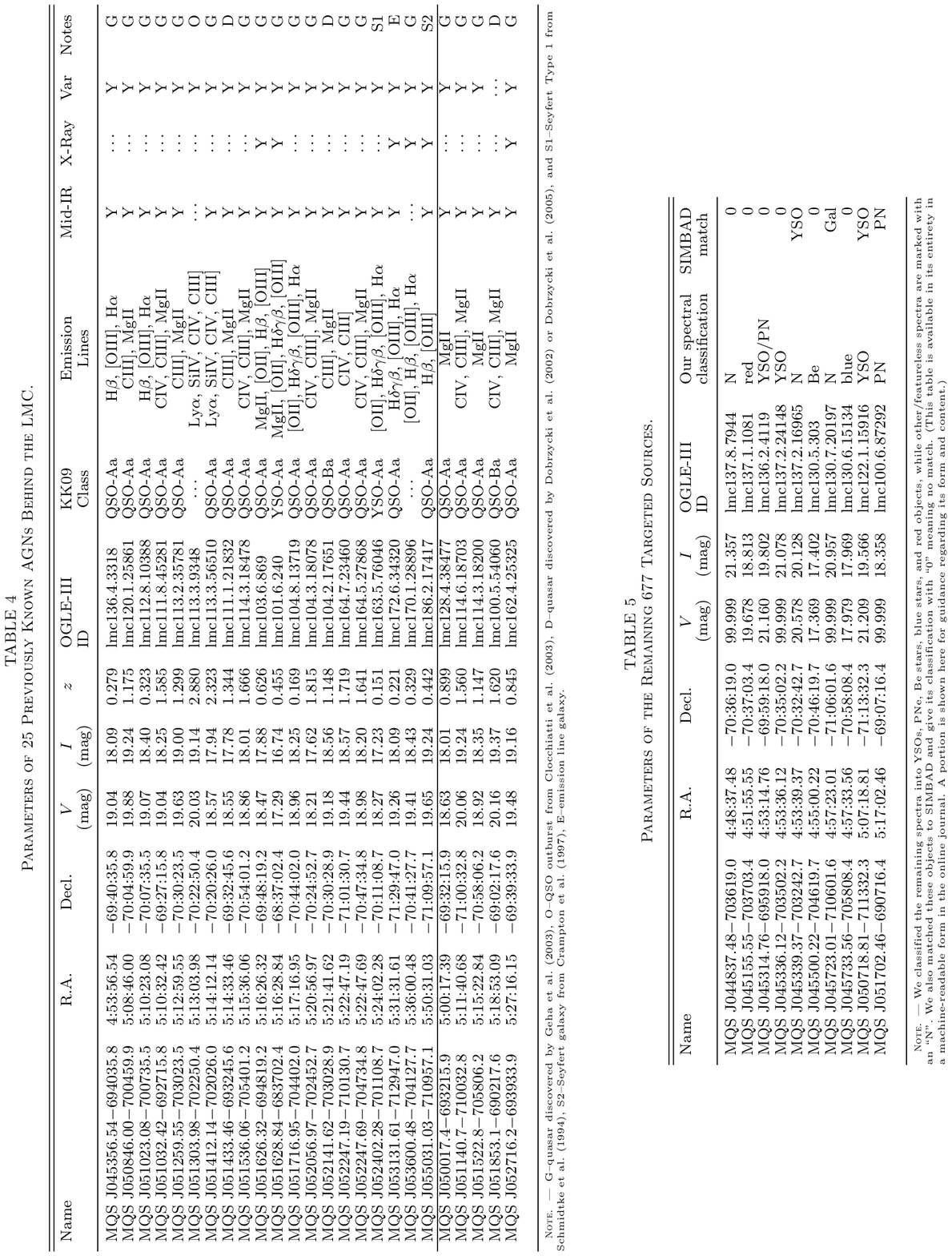}
\end{figure*}


\end{document}